\documentstyle[12pt]{article}

\begin{document}
\begin{center}
{\Large \bf Universe with the linear law of evolution }
\bigskip

{\large D.L.~Khokhlov}
\smallskip

{\it Sumy State University, R.-Korsakov St. 2, \\
Sumy 244007, Ukraine\\
E-mail: others@monolog.sumy.ua}
\end{center}

\begin{abstract}
The model of the homogenous and isotropic universe
is considered in which the coordinate system of reference
is not defined by the matter but is a priori specified.
The scale factor of the universe changes following
the linear law. The scale of mass changes proportional to
the scale factor. The temperature of the
relativistic matter changes inversly proportional to
the root of the scale factor. The model under consideration
avoids the flatness and horizon problems.
The predictions of the model are fitted to the observational
constraints: Hubble parameter, age of the universe,
magnitude-redshift relation of the high-redshift supernovae,
and primordial nucleosynthesis.
\end{abstract}

\section{Introduction}

As known~\cite{Dol,Lin}, the Friedmann
model of the universe has fundamental difficulties such as the
flatness and horizon problems due to the slow growth
of the scale factor of the universe.
In the Friedmann universe, the coordinate system of reference is
associated with the matter of the universe.
The evolution of the scale factor of the universe is given by
\begin{equation}
a \sim t^{1/2}, \qquad
a \sim t^{2/3}
\label{eq:a1}
\end{equation}
where the first equation corresponds to the matter as
a relativistic gas, and the second, to the dust-like matter.
Growth of the scale factor of the universe governed by
the power law with the exponent less than unity
is slower than growth of the horizon of the universe
$h \sim t$
that causes the flatness and horizon problems.

To resolve these problems an inflationary episode in the early
universe is introduced~\cite{Dol,Lin}. However there is another
way of resolving the problems. This is based on the premise
that the coordinate system of reference is not
defined by the matter but is a~priori specified.

\section{Theory}

Let us consider the model of the homogenous and isotropic
universe.
Let us assume that the coordinate system of reference is not
defined by the matter but is a~priori specified.
Let the coordinate system of reference be the Euclidean space
with the spatial metric $dl$ and absolute time $t$
\begin{equation}
dl^2=a(t)^2(dx^2+dy^2+dz^2), \quad t.
\label{eq:met}
\end{equation}
That is the coordinate system of reference is the
space and time of the Newton mechanics, with the scale factor of
the universe is a function of time.
Since the metric (\ref{eq:met}) is not defined by the matter,
we can a~priori specify the evolution law of the scale factor
of the universe. Let us take the linear law when
the scale factor of the universe grows with the velocity of light
\begin{equation}
a=ct.\label{eq:g1}
\end{equation}

In the Friedmann universe, the law~(\ref{eq:g1})
corresponds to the Milne model~\cite{Zeld}
which is derived from the condition that the density of the matter
tends to zero $\rho\rightarrow 0$. Here Eq.~(\ref{eq:g1})
describes the universe in which the evolution of the scale factor
do not depend on the presence of the matter. Hence the
density of the matter is not equal to zero $\rho\not= 0$.
The total mass of the universe relative to the background space
includes the mass of the matter and the energy of its gravity.
Let us adopt that the total mass of the universe is equal to zero,
that is
the mass of the matter is equal to the energy of its gravity
\begin{equation}
c^2={Gm\over{a}}.\label{eq:o}
\end{equation}
Allowing for Eq.~(\ref{eq:g1}),
from Eq.~(\ref{eq:o}) it follows that
the mass of the matter changes with time as
\begin{equation}
m={c^2a\over{G}}={c^3t\over{G}},\label{eq:p}
\end{equation}
and the density of the matter, as
\begin{equation}
\rho={{3c^2}\over{4\pi G a^2}}={3\over{4\pi G t^2}}.\label{eq:q}
\end{equation}
Hence the model under consideration yields the change of the scale
of mass proportional to the scale factor of the universe.
At the Planck time $t_{Pl}=(\hbar G/c^5)^{1/2}$,
the mass of the matter is equal to the Planck mass
$m_{Pl}=(\hbar c/G)^{1/2}$. At present, the mass of the matter is
of order of the modern value $m_0={c^2a_0/{G}}$.

Let us study time dependence of the temperature
of the relativistic matter.
Density of the relativistic matter is defined by its temperature
as
\begin{equation}
\rho\sim T^{4}.\label{eq:rht}
\end{equation}
From Eq.~(\ref{eq:q}) it follows that the temperature of the
relativistic matter changes with time as
\begin{equation}
T\sim a^{-1/2}\sim t^{-1/2}.\label{eq:tem}
\end{equation}

Let us consider the flatness and horizon problems within the
framework of the model under consideration.
Remind~\cite{Dol,Lin} that the horizon problem
in the Friedmann universe is caused by that the universe
observable at present
consisted of the causally unconnected regions in the past
that is inconsistent with the high isotropy of the background
radiation.
In the universe under consideration,
the size of the universe
(the scale factor of the universe)
coincides with the size of the horizon
during all the evolution of the universe.
Hence the presented model avoids the horizon problem.

Remind \cite{Dol,Lin} that
the essence of the flatness problem
in the Friedmann universe
is connected with impossibility to gain the
modern density of the matter at present starting from
the Planck density of the matter at the Planck time.
In the presented theory,
the density of the matter of the universe
changes from the Plankian value at the Planck time
to the modern value at the modern time.
Hence the flatness problem is absent in the presented theory.

\section{Predictions}

Write down some relations describing the universe via the
cosmological redshift given by
\begin{equation}
z=\frac{T}{T_{0}}-1\label{eq:red}
\end{equation}
where $T_{0}=2.73 \ {\rm K}$ is the modern
temperature of the cosmic microwave background.
In view of Eq.~(\ref{eq:tem}),
the scale factor of the universe at the redshift $z$ is given by
\begin{equation}
a(z)=\frac{a_{0}}{(z+1)^2},\label{eq:sf}
\end{equation}
and the age of the universe at the redshift $z$ is given by
\begin{equation}
t(z)=\frac{t_{0}}{(z+1)^2}.\label{eq:au}
\end{equation}
Angular diameter distance at the redshift $z$ is given by
\begin{equation}
d(z)=a_{0}\left(1-\frac{a}{a_{0}}\right)=
a_{0}\left(1-\frac{1}{(z+1)^2}\right).\label{eq:ad}
\end{equation}

Let us estimate the modern age of the universe from
Eq.~(\ref{eq:tem}). Taking into account change of
the electromagnetic constant $\alpha$,
we obtain
\begin{equation}
t_{0}=t_{Pl}\frac{\alpha_{0}}{\alpha_{Pl}}\left(\frac
{T_{Pl}}{T_{0}}\right)^2.
\label{eq:age}
\end{equation}
Allowing for that $\alpha_{0}=1/137$, $\alpha_{Pl}=1$,
$t_{PL}=5.39\cdot 10^{-44} \ {\rm s}$,
$T_{PL}=1.42\cdot 10^{32} \ {\rm K}$,
the modern age of the universe is equal to $t_{0}=
1.06\cdot 10^{18} \ {\rm s}=33.7 \ {\rm Gyr}$.
In view of Eq.~(\ref{eq:g1}),
the relation between the Hubble parameter and the age
of the universe is given by
\begin{equation}
H=\frac{1}{a}\frac{da}{dt}=\frac{1}{t}.\label{eq:j}
\end{equation}
Then the modern value of the Hubble parameter is $H_{0}=
0.94\cdot 10^{-18}\ {\rm s^{-1}}=29{\ \rm km/s/Mpc}$.

The observed Hubble parameter is derived from the relation
for the angular diameter distance.
In the Friedmann model,
the angular diameter distance at low redshifts is given by
\begin{equation}
d(z)\sim z\quad z\ll 1.\label{eq:hpr}
\end{equation}
In the presented model,
the angular diameter distance at low redshifts is given by
\begin{equation}
d(z)\sim 2z\quad z\ll 1.\label{eq:ad1}
\end{equation}
From this we must divide the observed Hubble parameter
obtained in the Friedmann model
by the factor 2.
If we adopt the observed Hubble parameter
obtained in the the Friedmann model
as $H_0=60\pm 10{\ \rm \ km/s/Mpc}$~\cite{Pat},
then in the presented model we obtain
$H_0=30{\rm \ km/s/Mpc}$ which is in agreement with
the above prediction.

Authors of~\cite{Steig} investigate
constraints on power-law models of the universe from the present
age of the universe, from the magnitude-redshift relation of
the high-redshift supernovae, and from
primordial nucleosynthesis.
Constraints from the current age of the universe~\cite{age}
and from the high-redshift supernovae data~\cite{Ia}
require $a\sim t$, while constraints from
primordial nucleosynthesis~\cite{Steig} require
$T^{-1}\leq t^{0.58}$.
Since in the
Friedmann universe $T^{-1}\sim a$, the above constraints are
inconsistent within the framework of the Friedmann universe.
The presented theory predicts $a\sim t$ and
$T^{-1}\sim t^{0.5}$ which are in agreement with
the above relations.

The age of the universe can be estimated by
the observed age of the oldest globular clusters.
Age of the globular cluster depends on
the distance to the cluster such that a revision of 0.1~mag
in the distance scale changes the age of the cluster by 10~\%.
Since the angular diameter distance
in the presented model is greater than that in the
Friedmann model by the factor 2,
the distance scale increases by $5lg2=1.5$~mag,
and the age of the oldest globular
clusters in the presented model is 2.5 times greater than
that in the Friedmann model.
If we adopt the observed age of the oldest globular clusters
obtained in the Friedmann model as
$t_{GC}=14\pm 2 {\ \rm Gyr}$~\cite{age},
then in the presented model, this yields
$t_{GC}=14\times 2.5=35{\ \rm Gyr}$ that is in agreement with
the age of the universe predicted by the presented model.

Deceleration parameter $q_{0}$ is defined via the
relation of luminosity distances at different redshifts.
The high-redshift supernovae data~\cite{Ia} favour
$q_{0}=0$.
In this case, the Friedmann and presented models
give the same relation of luminosity distances at different
redshifts.
So the estimate $q_{0}=0$ obtained from
the high-redshift
supernovae data within the framework of the Friedmann model
is also valid for the presented model.

Inequality $T^{-1}\leq t^{0.58}$~\cite{Steig}
is obtained from the condition that at the beginning of
nucleosynthesis when $T\sim 80 {\ \rm keV}$ the age of the
universe should be less than
the lifetime of neutron $t\leq 887 {\ \rm s}$.
The inferred primordial abundances of helium-4~\cite{helium}
and deuterium~\cite{deut} require $T^{-1}\sim t^{0.55}$~\cite{Steig}.
When $T^{-1} < t^{0.55}$,
deuterium is overproduced relative to helium-4.
Relation $T^{-1}\sim t^{0.55}$ corresponds to the modern age of the
universe $t_{0}=14 {\ \rm Gyr}$.
Increase of the modern age of the universe
up to $t_{0}=33.7 {\ \rm Gyr}$ gives $T^{-1}\sim t^{0.54}$.
From this relation $T^{-1}\sim t^{0.5}$ predicted by the
presented theory is inconsistent
with the inferred primordial abundances of helium-4 and deuterium.

\section*{Acknowledgements}

I thank G.~Steigman, M.~Kaplinghat, I.~Tkachev and T.P.~Walker
for drawing my attention to their paper astro-ph/9805114.

\end{document}